# *Analysis of the Effect of Atherosclerosis with the Changes of Hematocrit: A Computational Study on the Hemodynamics of Carotid Artery*

*Nihal Ahmed*
*Department of Material Science and Nanotechnology Bilkent University*
*Ankara, Turkey*
*nihalahmed129@gmail.com*

*Ashfaq Ahmed*
*Department of Mechanical and Aerospace Engineering*
*Oklahoma State University*
*Stillwater, Oklahoma, USA*
*ashfaq.ahmed@okstate.edu*

*Sakan Binte Imran*
*Sir Salimullah Medical College,*
*Dhaka – 1000, Bangladesh*
*sakanbinteimran.ssmc@gmail.com*

*Abstract*—**Atherosclerosis is a state in which plaque (fat, cholesterol, and other substances) builds up inside the arteries that eventually leads to carotid artery stenosis which is a stage of narrowing in the large arteries located on either side of the neck that carry blood to the brain, face, and head. Carotid stenosis is often associated with permanent injury of a part of the brain (strokes) due to loss of its blood supply. For example, ischemia mostly results in severe disability or death. Hematocrit or packed cell volume (PCV) is the volume of red blood corpuscles in relation to that of whole blood. The purpose of our research is to perform a Computational Fluid Dynamics (CFD) study of blood flow with the percentage changes of hematocrit to analyze the hemodynamic and physiological behavior of atherosclerosis. Our study a constructed 2D geometry model that has been analyzed using Finite volume method (FVM) for unique stages of atherosclerosis. The aim of this study is to explore the behavioral insights into the velocity gradient, wall shear stress, and pressure gradient of carotid artery under different percentages of hematocrit and different stages of atherosclerosis. The analysis presented conclusive difference in these parameters which can be used to identify the atherosclerosis formation.**

## I. INTRODUCTION

Atherosclerosis is an ailment of enormous and transitional measured veins wherein greasy injuries called atheromatous plaques creates within surfaces of the blood vessel dividers. For the most part, it alludes to thickened and solidified veins of all sizes [1] and it is liable for a fundamental reason for almost 50% of the apparent multitude of passings in westernized nations [2]. Looking at the age and probability of creating atherosclerosis, men are more inclined to endure than ladies which infer that man sex hormones may be atherogenic or, alternately that female sex hormones may be defensive. If there should be an occurrence of male, ordinary degree of hematocrit is 45% in normal (Range: 40-52%) and for female, it is 42% in normal (Range: 37-47%). To perform basic screening test for weakness viably [3] or to appraise the RBC and Hb check, the investigation of the level of hematocrit is vital. Expanded hematocrit brings about hemoconcentration because of any reason (consume, lack of hydration) and polycythemia because of any reason (polycythemia rubra vera, hypoxia) [4]. Interestingly, diminished hematocrit results hemodilution because of any reason (pregnancy or over the top liquid admission). As a component of circulatory framework, conduits convey the blood siphoned by the heart to the whole body while conveying oxygen and supplements to each cell. The information with respect to atherosclerosis permit us to forestall the potential inconveniences for example plaque developing inside the conduits. Respiratory failure may happen if the blood flexibly is decreased to the heart. A harmed heart muscle may not siphon also and can prompt cardiovascular breakdown. A stroke may happen if the blood flexibly is sliced off to the mind. Extreme agony and tissue passing may happen if the blood gracefully is decreased to the arms and legs.

Momen et al. [5] led a comparable report however the investigation of the level of hematocrit and its impact was at this point to be resolved. Christof et al. [6] showed the expected function of CFD in settling on choice with respect to treatment of vascular pathologies by zeroing in their investigation on the difference in divider shear worry within the sight of endoluminal unite. Jhunjhunwala et al. [7] examinations the non-Newtonian conduct of blood by exploring the speed and weight for different time steps. Padole et al. [8] utilizes the equivalent non-Newtonian and pulsatile conduct of blood to examine the adjustment in speed and weight and utilizing those change to figure the expanded circulatory strain. Greater part of the past examinations identified with atherosclerosis zeroed in on the adjustment in pressure angle and speed slope dependent on the phases of atherosclerosis. The thought of divider shear pressure was found in limited quantity of written works. That investigation was additionally founded on the different phases of atherosclerosis. In spite of the fact that the level of hematocrit assumes a major function in the improvement of coronary illness however past exploration work appears to overlook that parcel during investigation.

In this examination, the hemodynamics of the blood move through the carotid course was dissected for various rates of hematocrit (for example 25%, 45%, 60%) and various phases of atherosclerosis. The focal point of this investigation is to misuse the adjustments in the wall shear stress is dependent on the level of hematocrit. The examination likewise incorporates the thought of various phases of atherosclerosis. The arrangement follows a conventional methodology of considering blood stream as flimsy state and pulsatile stream. A blend of Navier Stokes condition and Bird Carreu model has been utilized for numerical demonstrating. Conclusive outcome shows an unmistakable connection between the wall shear stress and the level of hematocrit present in the blood.

## II. Physical Modeling

The analysis was done by using three physical two dimensional models of the carotid artery. Generally, the human artery consists of three different sections.

TABLE I. GEOMETRY INFORMATION OF THE MODEL

| **Components** | **Dimensions (mm)** |
|---|---|
| Central Carotid Artery | D = 5 |
| Internal Carotid Artery | 0.725D |
| External Carotid Artery | 0.63D |

The three sections consist of one central carotid artery which divides into two subsections called Internal Carotid Artery (ICA) and External Carotid Artery (ECA).

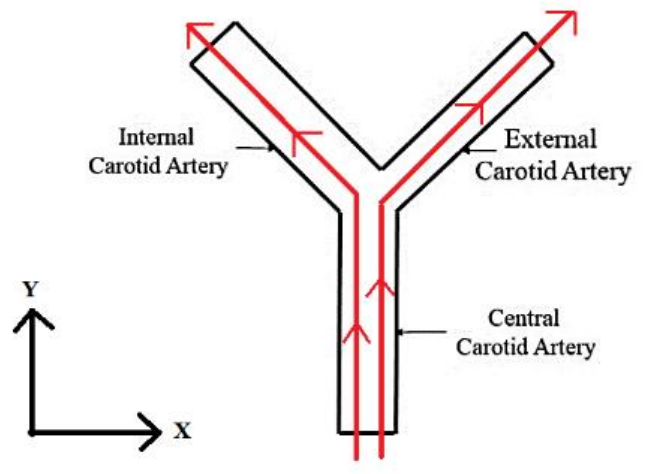

Fig. 1. Structure of a central cartid artery. (with blood flow direction)

The size of a central carotid artery varies from 5.3 mm to 6.9 mm [9]. Here the carotid bifurcation models were developed by the dimensions of Perktold [10] as shown in Table 1.

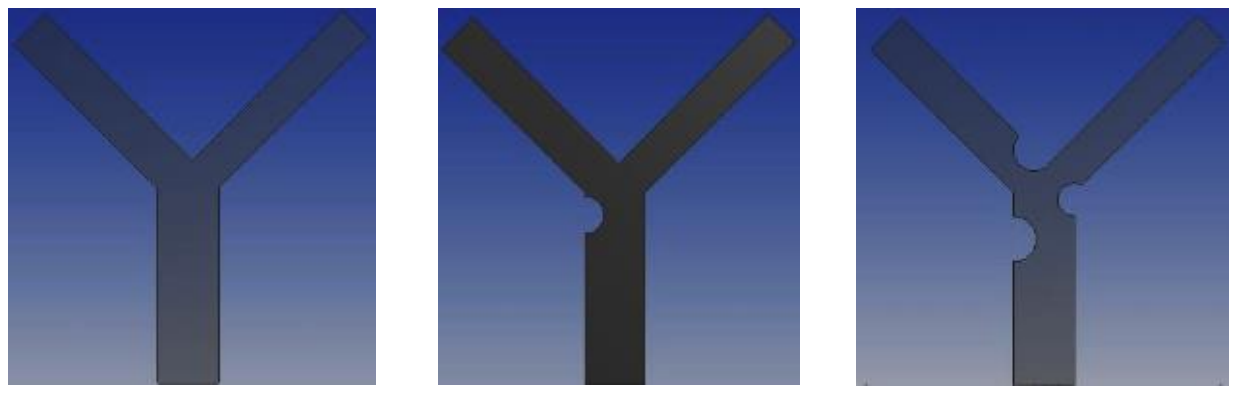

(a) (b) (c)

Fig. 2. 2D geometric model – (a) Healthy artery, (b) Initial state of atherosclerosis development and (c) Advanced state of atherosclerosis development.

Here, fig. 2 shows the structure of the three distinct models of carotid corridor that was utilized for our examination. The principal model alludes to a sound carotid supply route. The second and third models are the portrayal of the underlying plaque improvement and progressed plaque advancement which demonstrates the presence of atherosclerosis in the carotid supply route.

## III. Mathematical Modeling

### A. Governing equation and boundary condition

Blood can be considered as an incompressible liquid and blood vessel can be considered as deformable, rectilinear and isotropic material [11]. For taking care of this issue identified with incompressible liquid Navier Stokes condition has been recently utilized as a rule [12]. In our investigation, Navier Stokes condition and continuity condition have been utilized to portray the conduct of blood stream. Equation (1) and (2) are the Navier stokes condition and continuity condition that have been utilized for this investigation.

$$-\mu\nabla^2 u + \rho(u \cdot \nabla)u + \nabla p = 0 \quad (1)$$

$$\nabla u = 0 \quad (2)$$

Here P speaks to the pressure, ρ speaks to the blood density, μ is the image of blood kinematic viscosity. At the point when plaque fires developing in the conduit, it goes about as an unbending body. In this manner, it makes a hindrance in the stream. Therefore, blood streams from the distinction in absolute energy between two focuses. At the point when a vein is abruptly limited because of the presence of plaque, the velocity of the blood stream builds right the second it re-visitations of its unique distance across. Which eventually brings about the venturi pressure drop. This wonder is clarified utilizing Bernoulli's standard [13]. So as to describe the stream conduct with the changing weight the Bernoulli's condition was additionally utilized while planning the model. Equation (3) expresses the standard type of the Bernoulli's condition

$$P_1 + ½\rho u_1{}^2 + \rho g h_1 = P_2 + ½\rho u_2{}^2 + \rho g h_2 \quad (3)$$

Here P speaks to pressure, u speaks to blood speed, ρ speaks to blood speed and h speaks to stature. Our examination basically centers around the divider shear pressure and how does the adjustment in the level of hematocrit influence the divider shear pressure. Condition (4) speaks to the equation which was utilized for deciding the divider shear worry in the investigation. Here vein has been considered as a round and hollow chamber with unbending dividers. Beforehand so as to decide the vein divider shear pressure the Haagen-Poiseuille condition has been utilized [14]. So also, here shear stress for a blood stream experiencing a round and hollow divider, for example, vein has been communicated by utilizing the Haagen-Poiseuille equation.

$$\tau = 32.\mu.\frac{Q}{\pi.d^3} \quad (4)$$

Here, μ is the blood kinematic consistency, Q is the volumetric stream and d is the distance across of the vein. There has consistently been a discussion about blood being a Newtonian liquid or non-Newtonian liquid. Late investigations [15]–[17] recommend that blood transforms it consistency relying on the pressure applied to it and it additionally has shear diminishing properties. Accordingly, blood can be alluded to as a non-Newtonian liquid. Notwithstanding, non-Newtonian liquids can't be characterized a similar route as Newtonian liquid (for example water) is characterized. Blood doesn't have a fixed thickness. Along these lines, different models have been created for characterizing the non-Newtonian liquids, for example, power law model [18], Fledgling Carreu Model [19], Herschel-Bulkley Model [20]. In this investigation the Fowl Carreu model has been utilized to characterize the blood model. Condition (5) expresses the Fledgling Carreu Model. This model joins the Newtonian model of liquid and the force law model. This model can assess the conduct of liquid both under high shear rate and low shear rate. Subsequently this model was picked for this examination.

$$\mu = \mu_\infty + (\mu_0 - \mu_\infty)[1 + (\Gamma\ \dot{}\gamma)^2]^{\frac{n-1}{2}} \quad (5)$$

here, $\mu_\infty$= infinite shear viscosity, $\mu_0$ = zero shear viscosity, $n$ = power law index and $\gamma$ = time constant. For different hematocrits the parameters showed in Table 2 has been given as input.

TABLE II. BIRD CARREU MODEL PARAMETERS FOR DIFFERENT PERCENTAGES OF HEMATOCRIT [21].

| Hematocrit | Zero Shear Viscosity, $\mu_0$ | Infinite Shear Viscosity, $\mu_\infty$ | Power law index, n | Time constant, γ |
|---|---|---|---|---|
| 25% | 0.178 | 0.0257 | 0.330 | 12.448 |
| 45% | 1.610 | 0.0345 | 0.479 | 39.418 |
| 65% | 8.592 | 0.0802 | 0.389 | 103.088 |

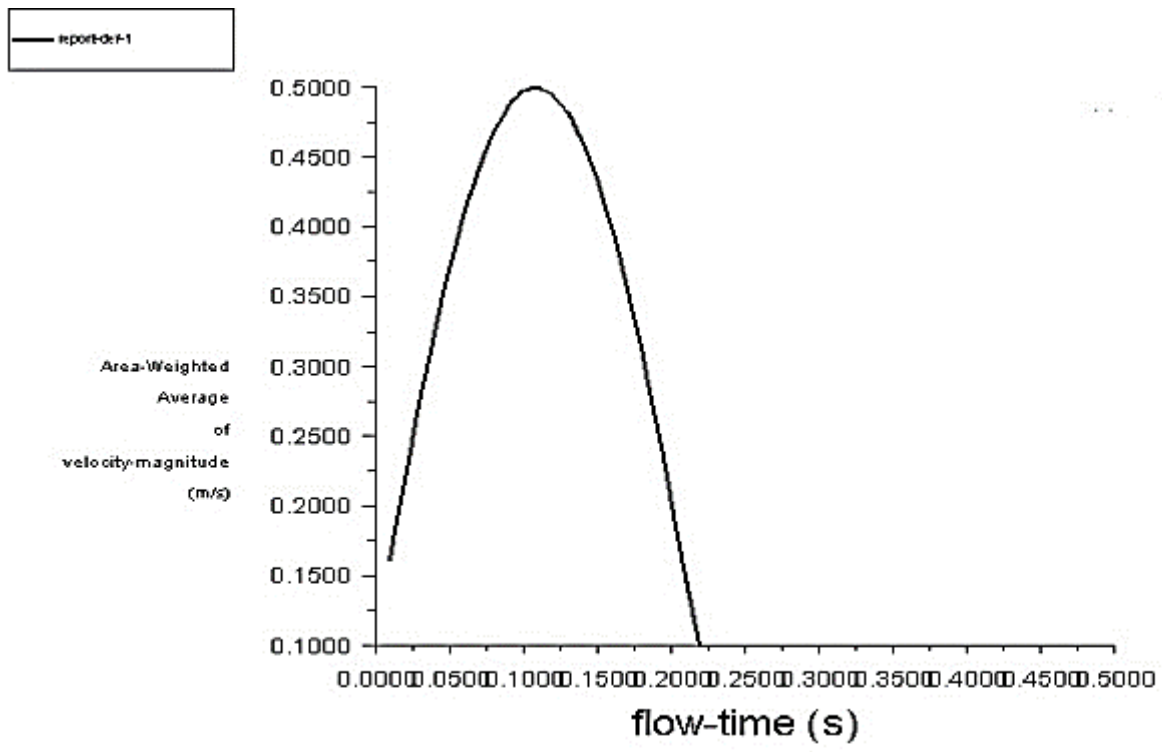

Fig. 3. Pulsatiel blood flow.

Blood flow is different from normal fluid flow. There is a factor named systolic and diastolic pressure associated with it. The artery flow is considered as a pulsatile flow.

TABLE III. BOUNDARY CONDITIONS

| Boundary | Applied Condition |
|---|---|
| Inlet | UDF pulsating velocity |
| Outlet | P = 13,332 Pa |
| Wall | No slip condition |

So as to communicate the stream, equation (6) has been utilized. Fig. 3 shows the throbbing stream that has been utilized in this investigation. Here at outlet a weight of 13,332 Dad was utilized which is the typical weight at the vein. Table 3 shows the total limit condition utilized in this examination.
The following equation represents the flow pattern at the boundary condition from this equation (6) -

$$v_{in}(t) = \begin{cases} 0.5\sin[4\pi(t + 0.0160236) & when\ 0.5n < x < 0.5n + .218 \\ 0.1 & when\ 0.5n + .218 < n < 0.5(n+1) \end{cases}$$

## IV. NUMERICAL MODELING

### A. Mesh sensitivity

While tackling any mathematical issue it is important to have excellent work. It decreases the odds of mistake and gives coherent answer for the conditions utilized in the arrangement strategy. Table 4 shows the quantity of lattice components utilized for this examination. In this examination, quadrilateral work was utilized toward the start. At that point further refinement close to the limit condition districts and plaque locales were utilized to support the exactness of the arrangement. At that point administering conditions were applied to singular hubs to locate the neighborhood parametric qualities. At that point these nearby qualities were joined to locate the worldwide qualities.

TABLE IV. NUMBER OFMESHING ELEMENTS USED IN THIS STUDY

| | Number of nodes | Number of elements |
|---|---|---|
| (a) | 26432 | 25837 |
| (b) | 24461 | 23866 |
| (c) | 23015 | 22414 |

While performing any numerical analysis it is essential that the results do not depend on the mesh element number. In fig. 4 the mesh sensitivity analysis for this solution has been shown and in fig. 5 we have shown the mesh region of carotid artery.

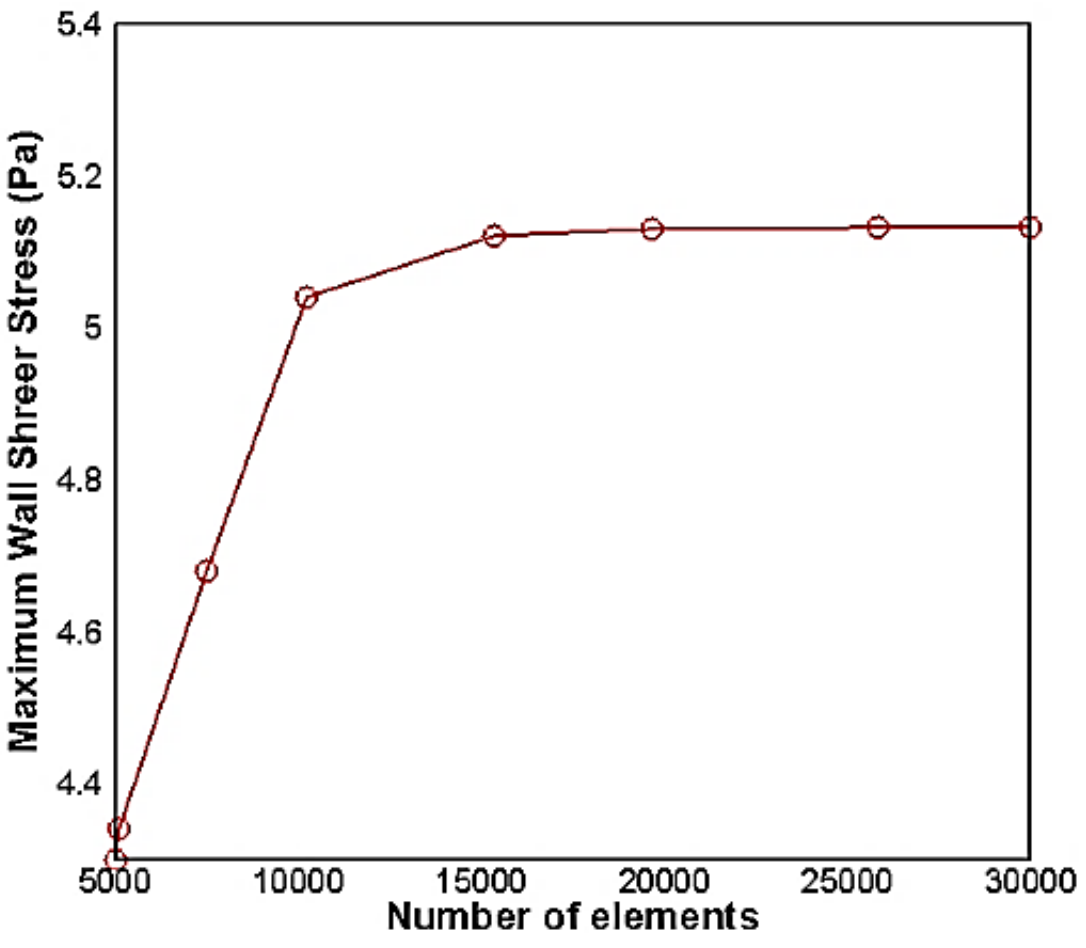

Fig. 4. Mesh sensitivity analysis.

The analysis suggests that the results were stable after twenty thousand elements. Considering the computational efficiency and solution accuracy around twenty-five thousand elements were used for the analysis.

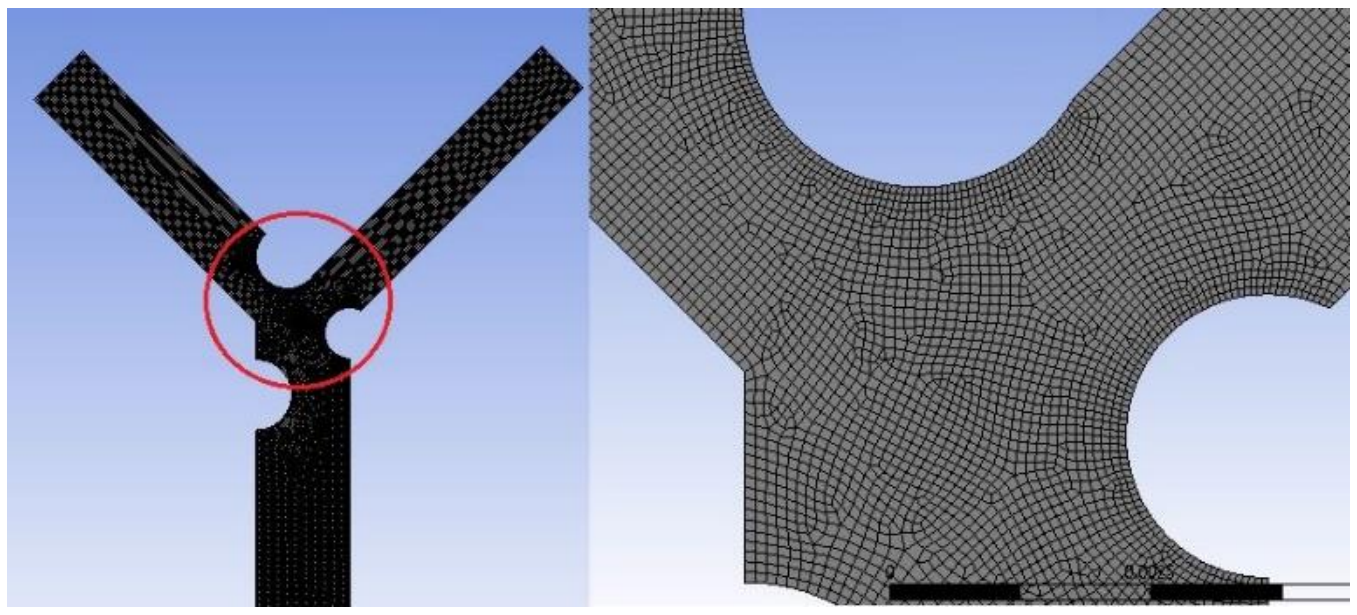

Fig. 5. Meshing region in carotid artery.

## V. NUMERICAL PROCEDURE AND VALIDATION

Normal artery flows are considered as laminar but unsteady flows [22]. Hence the problem in this paper has been considered as an unsteady state and all the simulations were done in unsteady state. Table 5 indicates the parameters associated with time that were used in this simulation.

TABLE V. UNSTEADY STATE PARAMETERS

| Conditions | Measurements |
|---|---|
| Time step size | 0.01s |
| Number of time steps | 50 |
| Maximum iteration repeating interval | 100 |

Fig. 6 shows the comparative analysis of wall shear stress with present study and the analysis of Momin et al. [5]. While the approach between the two studies were different but the outcome seems similar. The current numerical methodology was applied to the model geometry of the previous study and the wall shear stress behavior of both cases were similar.

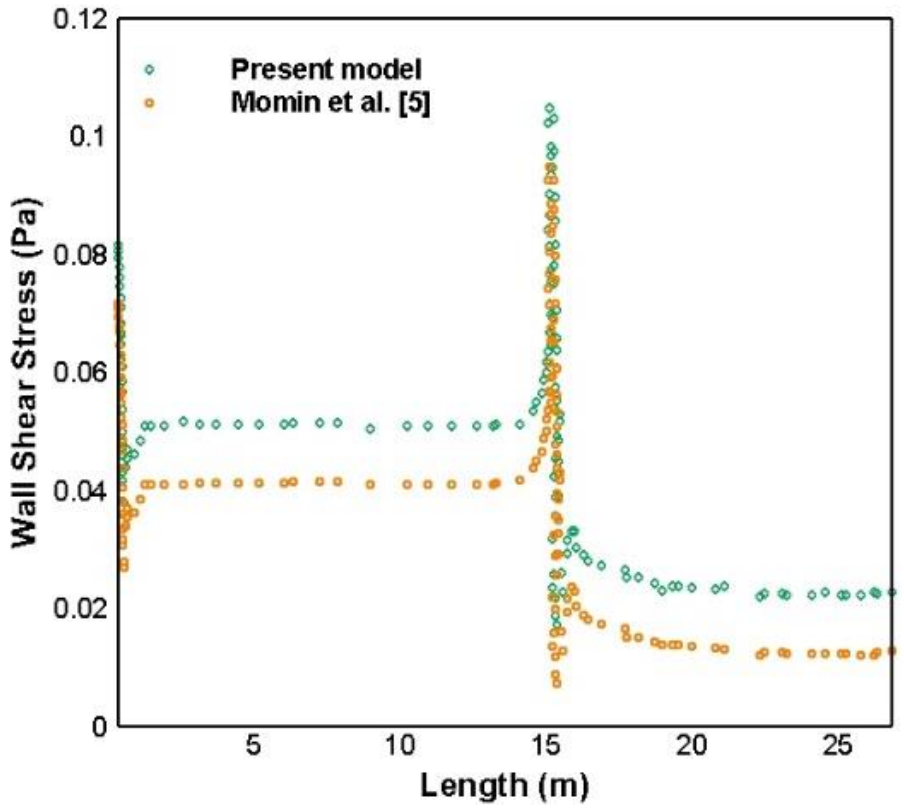

Fig. 6. Comparative analysis with previous studies.

There is a slight difference between the two models due to the difference in pulsatile flow property. Momin et al. [5] does not mention the complete profile of their inlet velocity. Hence some part of the velocity profile was assumed during programming the user defined function. That difference has resulted in a slight shift in the wall shear stress value. To verify our method, we further compared our results with studies with similar geometries. Jhunjhunwala et al. [7] used a similar Y shaped model to conduct his study to analyze the pulsatile and non-Newtonian behavior of blood flow. The value of maximum velocity for a healthy artery for different times are shown in table 6. The results are almost similar which indicates the accuracy of present approach.

| Time(s) | Present Study | Jhunjhunwala et al. [7] |
|---|---|---|
| 0.1 | 0.75 | 0.75 |
| 0.2 | 0.38 | 0.38 |
| 0.3 | 0.19 | 0.189 |

## VI. RESULTS AND DISCUSSIONS

The phenomenon talked about in this paper is unsteady henceforth the properties may change with time. It is fundamental to comprehend the conduct of different boundaries in understanding to time. Fig. 7 shows the adjustment in maximum wall shear stress with time. What's more, the investigation shows that the adjustment in wall shear stress stops after the stream has been reenacted for 0.13s. After 0.13s the stream gets steady because of its throbbing properties and after that stream properties become autonomous of time

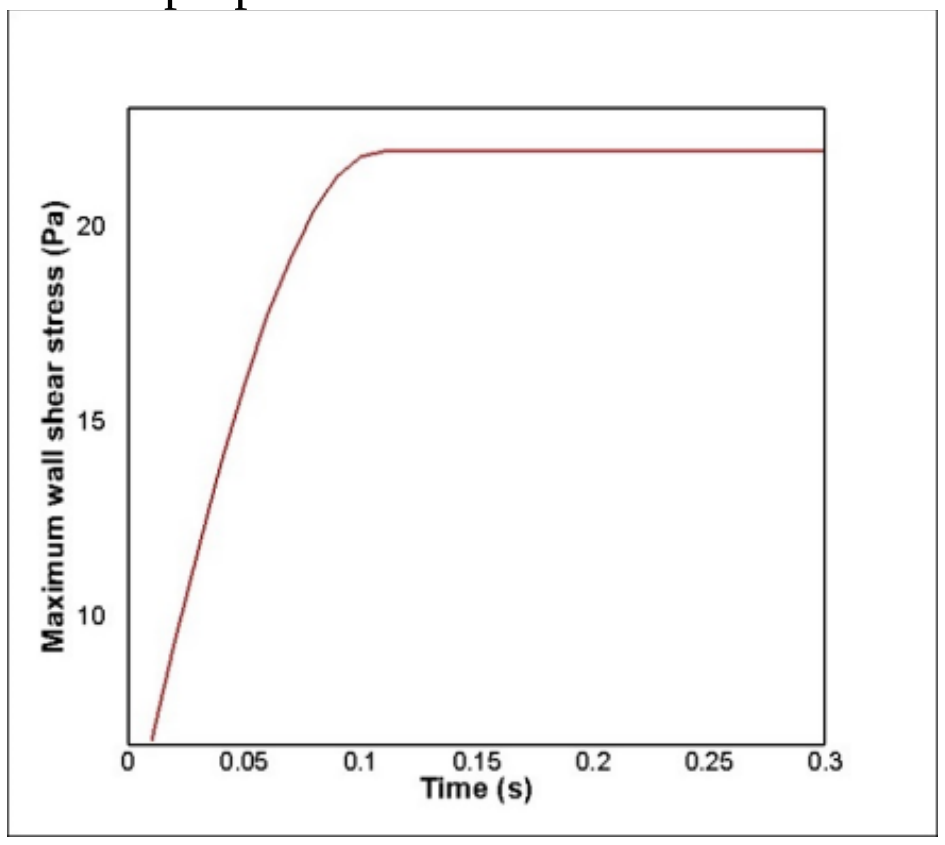

Fig. 7. Change in Maximum wall shear stress with time.

### A. *Changes in pressure and velocity gradient*

In this segment the relative investigation of the progressions of the pressure slope and velocity gradient with the blood stream were accomplished for various phases of atherosclerosis. Fig. 8 shows that for the solid conduit the weight is most extreme at the focal carotid corridor and least at the outside carotid course and the speed is least at the focal carotid vein and greatest at outer carotid supply route.

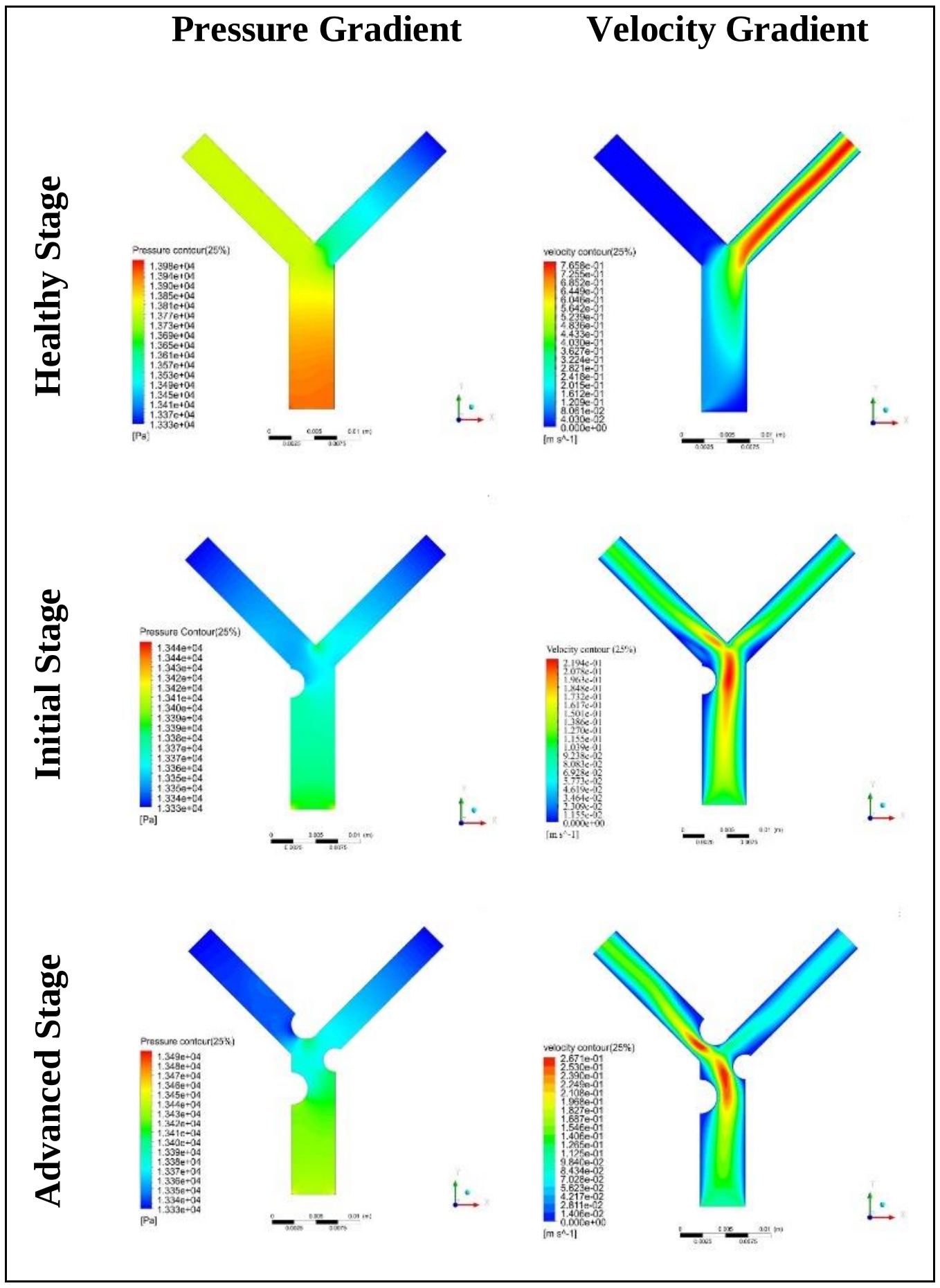

Fig. 8. Pressure and velocity gradient within the artery in three distinct stages of atherosclerosis at 0.5s (Hematocrit 25%).

This outcome can be defended by zeroing in on the Bernoulli's condition. As indicated by Bernoulli's condition, the pressure is most extreme where the zone is greatest. The velocity is greatest where zone is least.

The current outcome matches with the presumptions of Bernoulli on the grounds that the territory of focal carotid course is most extreme and the region of outer carotid vein is least. At the point when plaque begins creating at the underlying phases of atherosclerosis it brings about the decrease of territory of supply route. Which eventually brings about the pressure drop and an expansion in blood velocity.

Because of the expansion in blood course through the inner carotid vein which supplies blood to the brain. There is solid chance of a cerebrum stroke. Veins are at the degree of nano scale and consequently some of the time it gets hard to distinguish the presence of plaque utilizing the clinical instruments. Other than blood stream is characterized as a laminar stream however within the sight of plaque the stream is upset and slight aggravations can be seen in the stream design. These aggravations demonstrate some violent conduct of blood which can end up being deadly for patients in light of the fact that such disturbance can bring about overabundance pressure on the conduit and can prompt extreme heart ailments. Blood trades a few gases and supplements through the courses. Yet, because of an expansion in the blood flow, blood can't trade supplements appropriately which can prompt extreme entanglements in the physiology of the human body. Along these lines, it is fundamental to distinguish the adjustments in the elements of blood stream.

## B. *Changes in wall sheer stress with different perccentages of hematocrit*

This outcome can be advocated by zeroing in on the Bernoulli's condition. As per Bernoulli's condition, the pressure is most extreme where the region is greatest. The velocity is greatest where region is least.

Fig. 9 shows the adjustments in wall shear stress of the left focal carotid conduit wall with the adjustments in hematocrit. Red Platelet comprises of assortment of components, for example, Potassium, Calcium, and Magnesium. With an expansion in hematocrit, the volume of these components additionally increments in the blood. Accordingly, there is more contact in the middle of the blood components and artery walls. Thus, it results a noteworthy 182.35% expansion of the divider shear pressure. Fig. 9 additionally delineates that for three distinct rates of hematocrit, there is a basic raise in the divider shear pressure.

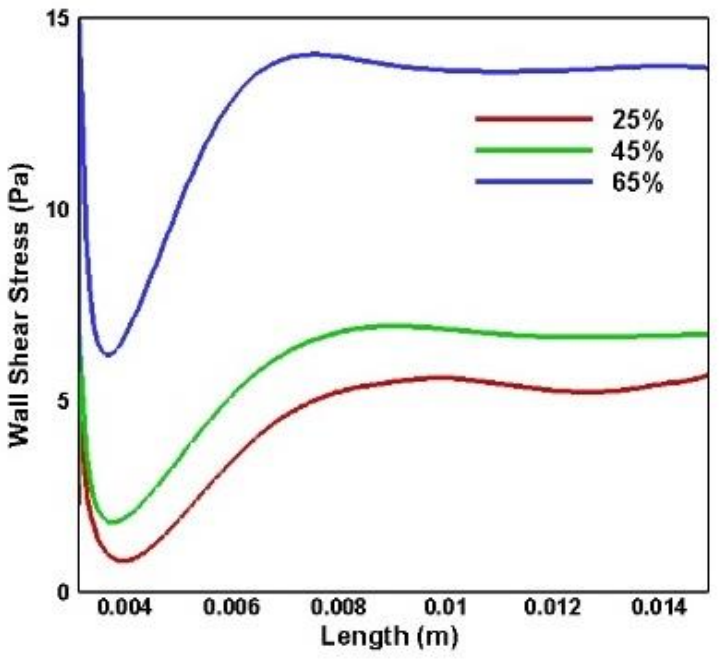

(a) Healthy Artery

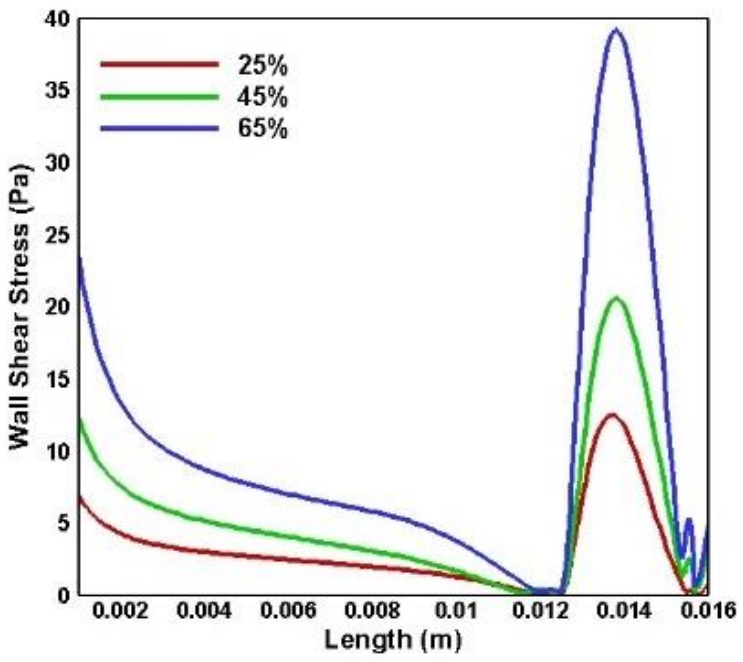

(b) Initial Stage of Atherosclerosis

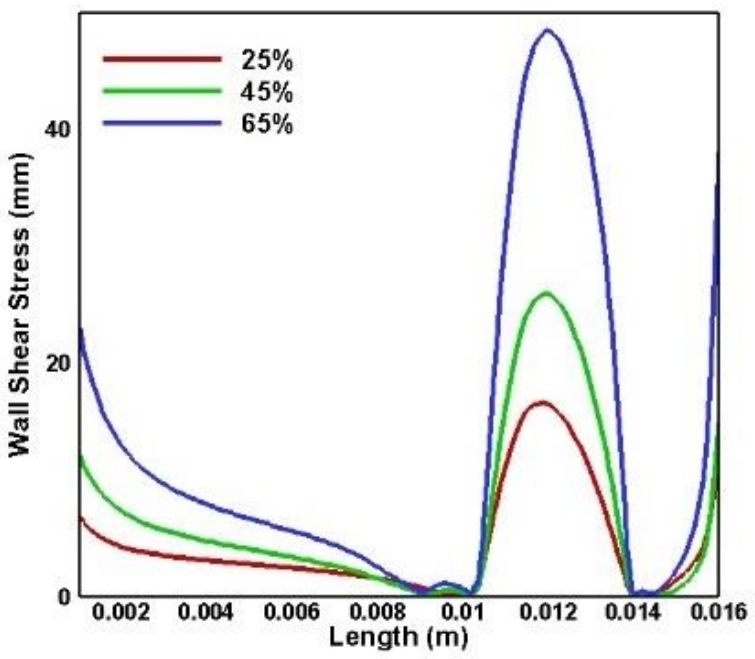

(c) Advanced Stage of Atherosclerosis

Fig. 9. Changes of wall shear stress (Pa) with respect to the changes in hematocrit under different stages of atherosclerosis formation.

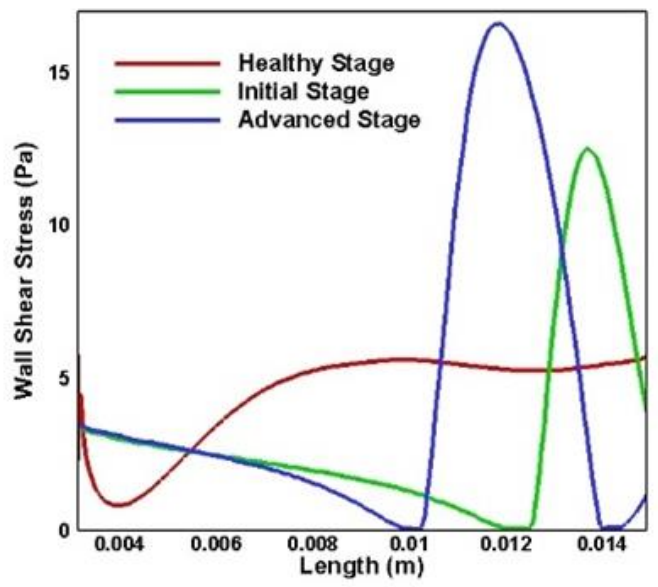

(a) Hematocrit (25%)

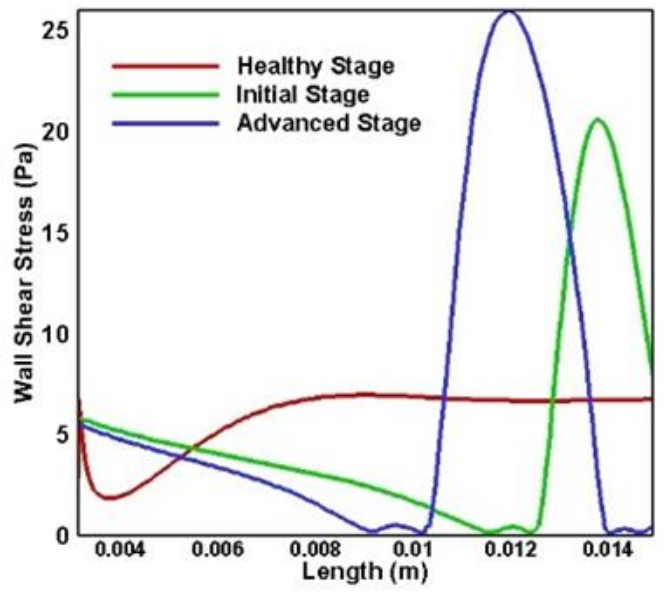

(b) Hematocrit (45%)

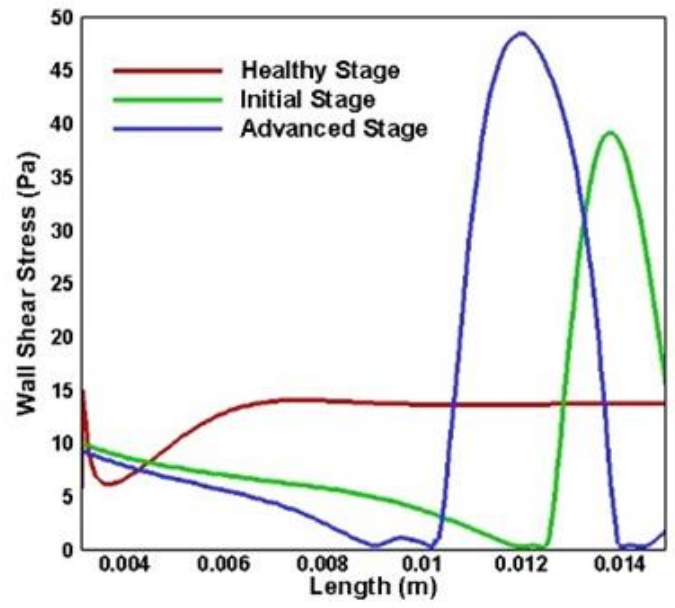

(c) Hematocrit (65%)

Fig. 10. Effect of the formation of atherosclerosis on developing wall sheer stress in a constant hematocrit level.

From fig. 10 we notice the adjustments in the wall shear stress of the left wall of the focal carotid artery. A noteworthy framing of shear stress at the plaque district is intense. Also, it is most extreme for the serious phases of atherosclerosis, which underpins our past investigation since the blood speed in that locale increments because of the decrease of zone. That expansion in blood speed at last outcomes in more friction in the vein wall lastly bringing about the expansion of wall shear stress. Aside from this, there is a steady conveyance of the wall shear stress in the sound supply route yet for starting and progressed stages the dispersion is unpredictable which can prompt other coronary course maladies.

## VII. CONCLUSION

The effect of atherosclerosis in focal vein was recreated in CFD and hemodynamic boundaries (for example pressure angle, wall shear stress and velocity inclination) were seen under various level of hematocrit accessible in blood. Result shows that the disturbance of blood stream and wall shear stress produced inside the conduit is unequivocally subject to the presence of hematocrit. Another significant finding from the velocity inclination is, the blood flow rate in the privilege carotid corridor is consistently more noteworthy than that of the left one since the left conduit contains atherosclerosis. Additionally, when plaque begins creating at the conduit in the extreme phase of atherosclerosis, there is more drop in the pressure and a bigger increment in the velocity. Because of the atherosclerosis, we have seen the blood stream changes to turbulent from laminar state. At last, from this investigation, we have a canny evaluation of the danger related with atherosclerosis as an element of the level of hematocrit in blood..

## REFERENCES

[1] Stevanovic, Nikola., "Guyton and Hall Textbook of Medical Physiology" - 13th Edition, pp. 872, 2009.

[2] S. Lewis Barbara Bain Imelda Bates, "Dacie and Lewis Practical Hematology", 10th Edition, pp. 31-33, 2006.

[3] Frank Firkin, C. Chesterman, D. Penington, B. Rush, "de Gruchy's Clinical Hematology in Medical Practice", 5th Edition, ISBN: 978-0-632-01715-7, pp. 24 – 27, 1989.

[4] "Hematology", *Health*, Johns Hopkins Medicine, 2015, https://www.hopkinsmedicine.org/health/treatment-tests-and-therapies/hematology

[5] Momin, M., Ara, N. and Arafat, M., "Assessment on The Effects of Atherosclerosis Formation in The Left Carotid Artery". AIP Publishing, Vol. 2121, Issue 1, DOI: 10.1063/1.5115934, 2019.

[6] Karmonik, C., Bismuth, J. X., Davies, M. G., & Lumsden, A. B., Computational hemodynamics in the human aorta: a computational fluid dynamics study of three cases with patient-specific geometries and inflow rates. Technology and Health Care, 16(5), 343-354, 2008.

[7] Jhunjhunwala, Pooja & Padole, Pramod & Thombre, S., CFD Analysis of Pulsatile Flow and Non-Newtonian Behavior of Blood in Arteries. Molecular & cellular biomechanics: MCB. 12. 37-47, 2015.

[8] Jhunjhunwala, Pooja & Padole, P & Thombre, S & Sane, Atul, Prediction of blood pressure and blood flow in stenosed renal arteries using CFD. IOP Conference Series: Materials Science and Engineering. 346. 012066, 2018. DOI: 10.1088/1757-899X/346/1/012066.

[9] J. Krejza et al., “Carotid artery diameter in men and women and the relation to body and neck size,” Stroke, vol. 37, no. 4, pp. 1103–1105, 2006.

[10] K. Perktold, M. Resch, and R. O. Peter, “Three-dimensional numerical analysis of pulsatile flow and wall shear stress in the carotid artery bifurcation,” J. Biomech., vol. 24, no. 6, pp. 409–420, 1991.

[11] J. Labadin and A. Ahmadi, “Mathematical modeling of the arterial blood flow,” in Proceedings of the 2nd IMT-GT Regional Conference on Mathematics, Statistics and Applications, Universiti Sains Malaysia, Penang, 2006.

[12] Saveljic, I., Nikolic, D., Milosevic, Z., Isailovic, V., Nikolic, M., Parodi, O., & Filipovic, N., 3D Modeling of Plaque Progression in the Human Coronary Artery. In Multidisciplinary Digital Publishing Institute Proceedings, Vol. 2, No. 8, p. 388, 2018.

[13] Klabunde, R. E.. Cardiovascular Physiology Concepts. Retrieved from https://www.cvphysiology.com/Hemodynamics/H012, 2020

[14] T. G. Papaioannou and C. Stefanadis, “Vascular wall shear stress: basic principles and methods,” Hell. J Cardiol, vol. 46, no. 1, pp. 9–15, 2005.

[15] H. E. A. Baieth, “Physical parameters of blood as a non-Newtonian fluid,” Int. J. Biomed. Sci. IJBS, vol. 4, no. 4, p. 323, 2008.

[16] F. J. H. Gijsen, F. N. van de Vosse, and J. D. Janssen, “The influence of the non-Newtonian properties of blood on the flow in large arteries: steady flow in a carotid bifurcation model,” J. Biomech., vol. 32, no. 6, pp. 601–608, 1999.

[17] F. J. H. Gijsen, E. Allanic, F. N. Van de Vosse, and J. D. Janssen, “The influence of the non-Newtonian properties of blood on the flow in large arteries: unsteady flow in a 90 curved tube,” J. Biomech., vol. 32, no. 7, pp. 705–713, 1999.

[18] A. Acrivos, M. J. Shah, and E. E. Petersen, “On the solution of the two-dimensional boundary-layer flow equations for a non-Newtonian power law fluid,” Chem. Eng. Sci., vol. 20, no. 2, pp. 101–105, 1965.

[19] J. L. Kokini, K. L. Bistany, and P. L. Mills, “Predicting steady shear and dynamic viscoelastic properties of guar and carrageenan using the Bird-Carreau constitutive model,” J. Food Sci., vol. 49, no. 6, pp. 1569–1572, 1984.

[20] I. H. Gucuyener, “A rheological model for drilling fluids and cement slurries,” in Middle East Oil Technical Conference and Exhibition, 1983.

[21] O. Kwon, M. Krishnamoorthy, Y. I. Cho, J. M. Sankovic, and R. K. Banerjee, “Effect of blood viscosity on oxygen transport in residual stenosed artery following angioplasty,” J. Biomech. Eng., vol. 130, no. 1, 2008.